\documentclass[aps,prl,twocolumn,superscriptaddress,notitlepage,nofootinbib]{revtex4-1}

\usepackage{epsfig}
\usepackage{multirow}
\usepackage{slashed}
\usepackage{amsmath}
\usepackage{physics}
\usepackage{qcircuit}
\usepackage{braket}
\usepackage{graphicx}
\usepackage{subfigure}
\usepackage{ulem}
\usepackage{color}
\usepackage{epstopdf}
\usepackage[inline]{enumitem}

\usepackage[breaklinks,colorlinks,urlcolor=blue,citecolor=citcolor,linkcolor=lcolor]{hyperref}
\definecolor{lcolor}{rgb}{0.5,0,0}
\definecolor{citcolor}{rgb}{0,0.3,0.0}
\definecolor{ao(english)}{rgb}{0.0, 0.5, 0.0}

\def\bea#1\eea{\begin{align}#1\end{align}}
\newcommand{\nn}{\nonumber\\}
\newcommand{\bef}{\begin{figure}[!htp]}
\newcommand{\eef}{\end{figure}}

\begin{document}

\title{Simulating Parton Fragmentation on Quantum Computers}
\author{Tianyin Li}
\affiliation{Key Laboratory of Atomic and Subatomic Structure and Quantum Control (MOE), Guangdong Basic Research Center of Excellence for Structure and Fundamental Interactions of Matter, Institute of Quantum Matter, South China Normal University, Guangzhou 510006, China}
\affiliation{Guangdong-Hong Kong Joint Laboratory of Quantum Matter, Guangdong Provincial Key Laboratory of Nuclear Science, Southern Nuclear Science Computing Center, South China Normal University, Guangzhou 510006, China}

\author{Hongxi Xing}
\email{hxing@m.scnu.edu.cn}
\affiliation{Key Laboratory of Atomic and Subatomic Structure and Quantum Control (MOE), Guangdong Basic Research Center of Excellence for Structure and Fundamental Interactions of Matter, Institute of Quantum Matter, South China Normal University, Guangzhou 510006, China}
\affiliation{Guangdong-Hong Kong Joint Laboratory of Quantum Matter, Guangdong Provincial Key Laboratory of Nuclear Science, Southern Nuclear Science Computing Center, South China Normal University, Guangzhou 510006, China}
\affiliation{Southern Center for Nuclear-Science Theory (SCNT), Institute of Modern Physics, Chinese Academy of Sciences, Huizhou 516000, China}

\author{Dan-Bo Zhang} 
\email{dbzhang@m.scnu.edu.cn}
\affiliation {Key Laboratory of Atomic and Subatomic Structure and Quantum Control (Ministry of Education), Guangdong Basic Research Center of Excellence for Structure and Fundamental Interactions of Matter, School of Physics, South China Normal University, Guangzhou 510006, China}

\affiliation{Guangdong Provincial Key Laboratory of Quantum Engineering and Quantum Materials,  Guangdong-Hong Kong Joint Laboratory of Quantum Matter, and Frontier Research Institute for Physics,\\  South China Normal University, Guangzhou 510006, China}

\collaboration{QuNu Collaboration}

\date{\today}         

\begin{abstract}
Parton fragmentation functions~(FFs) are indispensable for understanding processes of hadron production ubiquitously existing in high-energy collisions, but their first principle determination has never been realized due to the insurmountable difficulties in encoding their operator definition using traditional lattice methodology. We propose a framework that makes a first step for evaluating FFs utilizing quantum computing methodology. The key element is to construct a semi-inclusive hadron operator for filtering out hadrons of desired types in a collection of particles encoded in the quantum state. We illustrate the framework by elaborating on the Nambu-Jona-Lasinio model with numeral simulations. Remarkably, We show that the semi-inclusive hadron operator can be constructed efficiently with a variational quantum algorithm. Moreover, we develop error mitigation techniques tailed for accurately calculating the FFs in the presence of quantum noises. Our work opens a new avenue for investigating QCD hadronization on near-term quantum computers.
\end{abstract}

\maketitle

\vspace{0.5cm}
\noindent{\it Introduction}.- Parton fragmentation functions~(FFs) quantify the probability densities for the transition of colored partons to colorless hadrons~\cite{Berman:1971xz,Feynman:1972,Collins:2011zzd}. It stands out as an essential ingredient for describing high energy hadron production in electron-positron, lepton-hadron, and hadron-hadron collisions \cite{Metz:2016swz}. Parton FFs, as dynamical non-perturbative quantities, not only can help us to understand the fundamental mechanism of QCD hadronization, but also provide us with a crucial tool to access the nucleon structure in the upcoming era of electron-ion colliders \cite{AbdulKhalek:2021gbh,Anderle:2021wcy}, and to probe the transport properties of quark gluon plasma created in heavy ion collisions at Relativistic Heavy Ion Collider and the Large Hadron Collider \cite{JET:2013cls,Qin:2015srf}. 

Determination of FFs from the fundamental QCD theory is crucial for our understanding of color confinement. In the past decades, tremendous efforts have been devoted for extracting FFs from global analysis of world data based on the QCD factorization theorem, see e.g. ~\cite{deFlorian:2007aj,Hirai:2007cx,Bertone:2018ecm,AbdulKhalek:2022laj,Soleymaninia:2022alt,Moffat:2021dji,Li:2024etc,Gao:2024nkz,Borsa:2022vvp}. However, direct calculation of FFs in a first-principle fashion is still in lack of exploration, in sharp contrast with their counter part nonperturbative functions like parton distribution functions (PDFs) and light cone distribution amplitudes, where the relevant lattice investigation has formed a main branch in particle and nuclear physics community \cite{Lin:2017snn}. The main challenges that hinder the lattice calculation of FFs are mainly due to the following two intrinsic difficulties. One reason is that FFs are essentially real-time dynamical quantities, their direct calculation using lattice QCD encounters sign problems in the traditional Monte Carlo method~\cite{Alexandru:2016gsd}. In addition, in the operator definition of FFs, the key point is to construct the semi-inclusive hadron operator (SIHO) $\mathcal{H}\equiv\sum_X\ket{h,X}\bra{h,X}$, in which one needs to sum over all possible multi-particle states $\ket{h,X}$ containing the identified hadron $h$. However, the unidentified $X$ can be any state as long as they satisfy the QCD conservation laws \cite{Collins:2023cuo}. Such fickle states significantly challenge the encoding of $X$ in lattice QCD framework, and the sum over of all possible $X$ involves an exponentially increasing complexity in the numeral calculation. Addressing these difficulties lies far beyond the capabilities of classical computing for FFs in a first principle fashion.

Quantum computing, as a promising method alternative to the classical Monte Carlo method, has attracted lots of attention to simulate generic quantum systems~\cite{Feynman:1981tf} involving QCD. Typical examples including the proof of polynomial time for simulating quantum field theory using quantum computing \cite{Jordan:2012xnu}, followed by tremendous ideas for the evaluation of real time dynamics of QCD, such as the calculation of hadron partonic structure~\cite{Lamm:2019uyc,Echevarria:2020wct,Mueller:2019qqj,Qian:2021jxp,Li:2021kcs,Li:2022lyt,Grieninger:2024cdl}, simulation of parton shower~\cite{Bauer:2019qxa,Bepari:2021kwv,Bauer:2023ujy}, jet physics ~\cite{Florio:2023dke,Qian:2023vzn,Barata:2022wim} and string breading in thermal medium \cite{Lee:2023urk,Xie:2022jgj}, as well as the calculation of scattering amplitude~\cite{Briceno:2020rar,Li:2023kex,Barata:2024bzk}. For more broad applications, see e.g. Refs. \cite{Bauer:2022hpo,Bauer:2023qgm}. Inspired by the great advantage of using quantum computing for QCD real time problems, it is foreseeable to extend the quantum computing approach, especially the one for direct calculation of PDFs from light-cone correlators~\cite{Li:2022lyt}, for evaluating FFs. Nevertheless, how to formulate a semi-inclusive process and efficiently encode the associated hadron operator with qubits remains unexplored for quantum computing. 

In this letter, we propose a quantum computing approach for the first direct evaluation of parton fragmentation functions. We present a general framework to construct the semi-inclusive hadron operator for filtering out the identified hadrons in a collection of particles encoded in the quantum state. Utilizing a variational quantum algorithm, FFs are efficiently converted to multi-point dynamical correlators. We validate the approach in a prototypical model of Nambu-Jona Lasinio (NJL) using classical hardware~\cite{Nambu:1961fr,Nambu:1961tp,Gross:1974jv}. Moreover, we develop error mitigation techniques tailed for accurately calculating the FFs in the presence of quantum noises. Our work clarifies the operational meaning of the semi-inclusive process as well as provides a first step for calculating FFs on near-term quantum processors. 

\vspace{0.5cm}
\noindent{\it Encoding FFs on quantum computer}.- The operator definition of collinear quark fragmentation function in $d$-dimension spacetime can be written as \cite{Collins:2011zzd}
\bea
\label{eq:ff-def}
D^h_q(z) =&z^{d-3}\int \frac{dy^-}{4\pi} e^{-i y^- p^+/z} \Tr\Big\{\bra{\Omega} \psi(y^-)\nn
	&\times\sum_X\ket{h,X}
	\bra{h,X}\bar{\psi}(0) \ket{\Omega}\gamma^+\Big\} \,,
\eea
where $z$ is the momentum fraction of the fragmenting quark $q$ carried by the identified hadron $h$. In Eq. (\ref{eq:ff-def}), the Wilson line is implicit, $\psi(y^-) = e^{iHt}\psi(-y^z)e^{-iHt}$ is quark field on light-cone with light-cone coordinates defined as $y^{\pm}=(t\pm y^z)/\sqrt{2}$. $\ket{\Omega}$ is the vacuum state and $p^+$ is the light cone momentum of hadron $h$. $\ket{h,X}$ denotes a state that encodes hadron $h$ with momentum $p^+$ as well as $X$ representing additional unidentified fragments. The summation of $X$ in Eq.~\eqref{eq:ff-def} stresses that only the hadron $h$ is observed. 

\begin{figure*}[htbp]
	\centering
        \includegraphics[width=0.8 \textwidth]{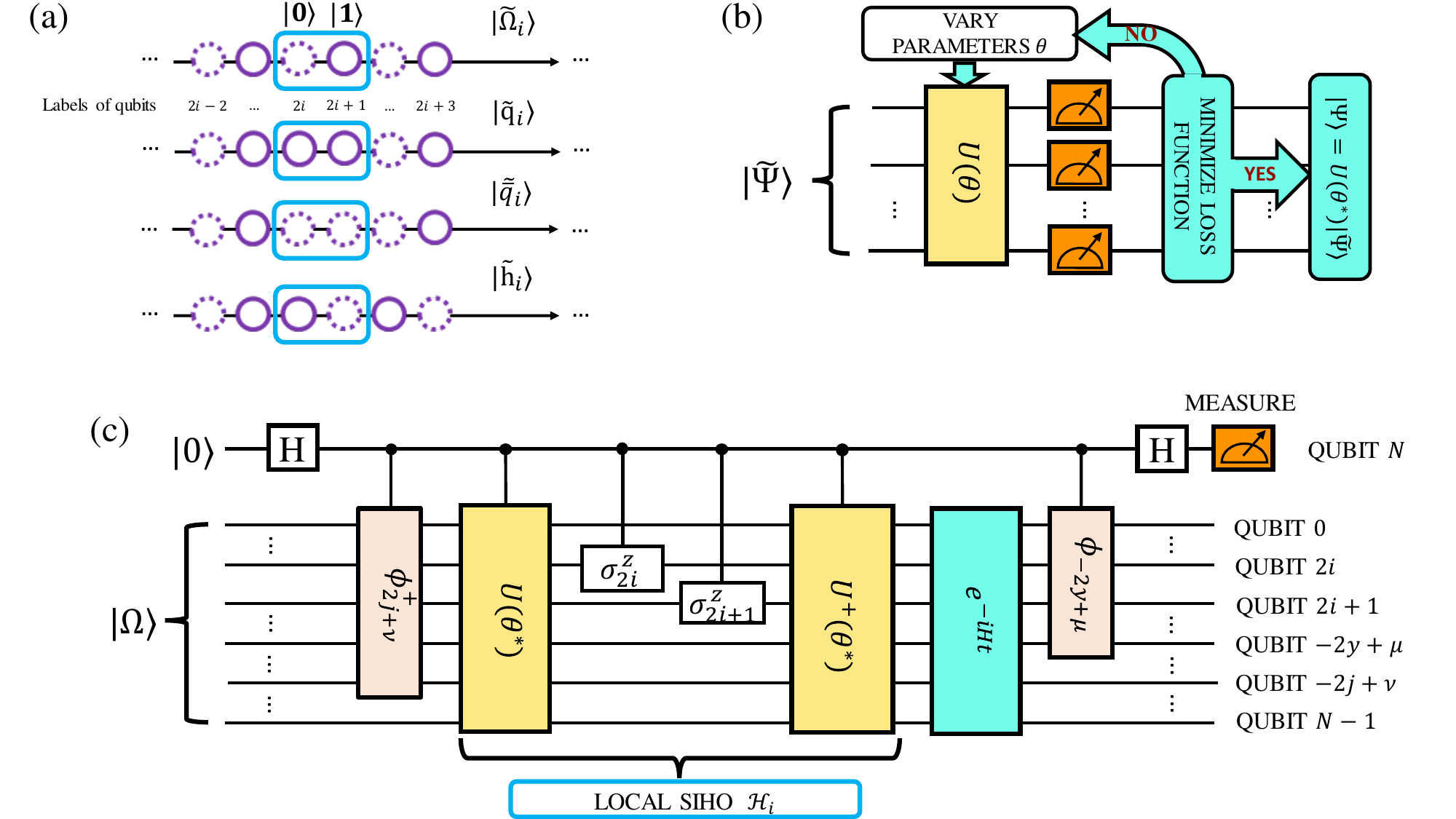}        \caption{
        Figure (a) shows qubit configurations of bare vacuum and one-particle states of in (1+1)-dimensional one flavor theory.
        (b) is the quantum circuit for variational quantum algorithms to prepare the vacuum and one-particle state.
        (c) is the quantum circuit for the simulation of Eq.~(\ref{eq:posFF}). }
	\label{fig-qcircuit}
\end{figure*}

To address the challenge for direct evaluation of FFs, it is inspiring to compare it with PDFs. PDFs involve dynamical light-cone correlator $\bra{p}\bar\psi(y^-)\gamma^+\psi(0)\ket{p}$, which can be calculated using the quantum algorithm as elaborated in our previous work~\cite{Li:2021kcs}. In addition to this dynamical light-cone correlator, the semi-inclusive hadron operator $\mathcal{H}\equiv\sum_X\ket{h,X}\bra{h,X}$ is inserted in the definition of FFs. The most challenging part of simulating FFs is the construction of $\mathcal{H}$ due to the existence of unidentified particles $X$. For quantum computing, figuring out concisely what are $h$ and $X$ in $\ket{h,X}$ as well as construction of $\mathcal{H}$ is unexplored. The goal of this work is to make the first step both in concept and in practice on how to construct the semi-inclusive hadron operator $\mathcal{H}$ and perform a direct evaluation of parotn FFs.

We propose a scheme for constructing $\mathcal{H}$ in a general setting. In particular, we present the methodology in $(1+1)$-dimensional system including the preparation of multi-particle states and the construction of $\mathcal{H}$, the extension to $(3+1)$-dimensional case can be pursued in a similar fashion. 

To be general, we introduce a computational basis $\ket{\boldsymbol{I}}$ of quantum computing for a one-dimensional spacial lattice system with total space site $M$, and each site contains $K$ qubits, i.e. $\ket{\boldsymbol{I}}=\ket{\boldsymbol{I}_0,\boldsymbol{I}_1,...,\boldsymbol{I}_{M-1}}$. $\boldsymbol{I}_{i}$ is a bitstring of $K$ bits encoding the internal degrees of freedom, such as spin, flavor, and color, in addition to the stagger fermion. The total number of qubits for the lattice system is $N=KM$.  For each $\boldsymbol{I}_{i}$, $2^K$ configurations can be mapped to bare vacuum ($\boldsymbol{I}^\Omega_i$), different types of bare elementary particles ($\boldsymbol{I}^q_i,\boldsymbol{I}^{\bar{q}}_i,\boldsymbol{I}^g_i,...$ for quark, anti-quark and gluon,...), and different types of bare hadrons ($\boldsymbol{I}^h_i$ with $h=\pi^\pm, K^\pm,...$). The bare particles are defined in the large quark mass limit $m\to \infty$. For example, as shown in Fig.~\ref{fig-qcircuit} (a), the bare vacuum, bare quark $q$, bare anti-quark $\bar{q}$ and bare hadron $h_{q\bar{q}}$ are $\ket{\boldsymbol{I}^\Omega_i}=\ket{01}$, $\ket{\boldsymbol{I}^q_i}=\ket{11}$, $\ket{\boldsymbol{I}^{\bar{q}}_i}=\ket{00}$ and $\ket{\boldsymbol{I}^{h_{q\bar{q}}}_i}=\ket{10}$ in the $(1+1)$-dimensional one flavor theory. Then, one can convert them to physical vacuum $\ket{\Omega}$ and physical single-particle states $\ket{\Psi}$ with a unitary transformation $U$ that preserve the quantum numbers
\bea\label{eq:U}
\ket{\Omega}= U\ket{\Tilde{\Omega}}, 
~~~\ket{\Psi}=\frac{1}{\sqrt{M}} \sum_{i=0}^{M-1}U\ket{\Tilde{\Psi}_i},
\eea
where $\ket{\Tilde{\Omega}} \equiv \ket{\boldsymbol{I}^\Omega_0,\boldsymbol{I}^{\Omega}_1,...,\boldsymbol{I}^\Omega_{M-1}}$, $\ket{\Tilde{\Psi}_i}\equiv\ket{\boldsymbol{I}^\Omega_0,\boldsymbol{I}^{\Omega}_1,...,\boldsymbol{I}^{\Omega}_{i-1},\boldsymbol{I}^{\Psi}_{i},\boldsymbol{I}^{\Omega}_{i+1},...,\boldsymbol{I}^{\Omega}_{M-1}}$ is the bare one-particle state at space point $i$, and the wavepacket of a physical particle $\ket{\Psi}$ centered in the $i$-th space point can be expressed as $\ket{\Psi_i}=U\ket{\Tilde{\Psi}_i}$. Notice that $\Psi$ stands for all one-particle configurations $\Psi=q,\bar{q},g,...,h$. In this setup, a state with two particles $\ket{\Psi}$ and $\ket{\Psi'}$ centered in the $i$-th and $j$-th space point can be prepared as
\bea\label{eq:twh}
    \ket{\Psi_i,\Psi'_j}= U\ket{\Tilde{\Psi}_i,\Tilde{\Psi}'_j}\,,
\eea
where $\ket{\Tilde{\Psi}_i,\Tilde{\Psi}'_j}\equiv\ket{\boldsymbol{I}^\Omega_0,...,\boldsymbol{I}^\Omega_{i-1},\boldsymbol{I}^{{\Psi}}_{i},\boldsymbol{I}^\Omega_{i+1}...,\boldsymbol{I}^{{\Psi}'}_{j},...,\boldsymbol{I}^\Omega_{M-1}}$ is the bare two-particle state.
Likewise, the multi-particle state can also be prepared by acting $U$ on the bare multi-particle state $\ket{\Tilde{\Psi}_i,\Tilde{\Psi}'_j,\Tilde{\Psi}''_k,...}$, which is also the computational basis. The definition of the bare multi-particle state is similar to the bare two-particle state.

Now, we are ready to define $X$ on a quantum computer. $X$ contains vacuum and/or all possible particles, i.e. $X = \Omega,q,\bar{q},g,...,h$, as long as the quantum number of the fragmenting parton is conserved in transition to $h+X$ system. The $\ket{h,X}$ state with $h$ centered in space point $i$ and $X$ distributed in the space point except $i$ can be written as

\bea
\ket{h_i,X_{\{j\not=i\}}}= U\ket{\Tilde{h}_i,\Tilde{X}_{\{j\not=i\}}}\,,
\eea
where $\ket{\Tilde{h}_i,\Tilde{X}_{\{j\not=i\}}} \equiv \ket{\boldsymbol{I}^{X_0}_0,...,\boldsymbol{I}^{X_{i-1}}_{i-1},\boldsymbol{I}^h_{i},\boldsymbol{I}^{X_{i+1}}_{i+1},...,\boldsymbol{I}^{X_{M-1}}_{M-1}}$. 
Then the local operator $\mathcal{H}_i$ can be constructed as
\bea
    \mathcal{H}_i
    &=U \Tr_{\{j\not= i\}} \ket{\Tilde{h}_i,\Tilde{X}_{\{j\not=i\}}} \bra{\Tilde{h}_i,\Tilde{X}_{\{j\not=i\}}} U^\dagger\nn
    &=U\ket{\boldsymbol{I}^h_i}\bra{\boldsymbol{I}^h_i}\otimes {\rm Id}_{\{j\not= i\}}U^\dagger\,,
\eea
where $\Tr_{\{j\not=i\}}=\sum_{X_0,...,X_{i-1},X_{i+1},...X_{M-1}}$ stands for the summation over all possible $X$. Thanks to the direct product structure and the completeness of computational basis $\ket{\Tilde{h}_i,\Tilde{X}_{\{j\not=i\}}}$, the trace leads to identity operator ${\rm Id}_{\{j\not= i\}}$ that acts on all qubits except for the space point $i$. For the sake of convenience and considering that FFs are Lorentz invariant, we perform our calculation in the hadron rest frame. 
To select out the zero-momentum hadron state, we construct the zero-momentum projector $\frac{1}{M}\sum_{j=0}^{M-1} T^j$, where $T$ is the translation operator that commutes with $U$. Therefore, the semi-inclusive hadron operator in FFs can be written as 
\bea\label{eq:proqu}
 \mathcal{H}=\left(\sum_{i=0}^{M-1}U\ket{\boldsymbol{I}^h_i}\bra{\boldsymbol{I}^h_i}U^\dagger\right)\times \left(\frac{1}{M}\sum_{j=0}^{M-1}T^j\right)\,,
\eea
where the operator ${\rm Id}_{\{j\not= i\}}$ is implicit, and more details about $\mathcal{H}$ can be found in Sec. A of supplemental material.

The next key ingredient is to construct the unitary matrix $U$ that transforms the computational basis of bare particles into states of physical particles. As shown in Fig.~\ref{fig-qcircuit} (b), to realize $U$ on a quantum computer, we refer to a variational quantum computing approach, where the unitary matrix is parameterized as $U(\theta)$ with a Hamiltonian variational ansatz which can preserve the quantum numbers~\cite{Li:2021kcs}, see details in Sec. B of supplemental material. Correspondingly, vacuum trial states $\ket{\Omega(\theta)}$, one-particle trial states $\ket{\Psi(\theta)}$ and two-particle trial states $\ket{\Psi(\theta),\Psi'(\theta)}$ can be generated by replacing $U$ as $U(\theta)$ in Eqs.~\eqref{eq:U} and~\eqref{eq:twh}. The generation of multi-particle trial states can be pursued similarly. To optimize the parameters $\theta$, we implement the subspace-search VQE algorithm by minimizing a weighted summation of variational energies of $\ket{\Omega(\theta)}$ and different multi-particle trial states $\ket{(\Psi,\Psi',\Psi'',...)(\theta)}$~\cite{PhysRevResearch.1.033062}. In particular, considering limited computing resources, we apply a truncation of the multi-particle states. We argue that it is a reasonable approximation to only generate the vacuum trial state $\ket{\Omega(\theta)}$ and one-particle trial state $\ket{\Psi(\theta)}$ if the hadron-hadron interaction energy $V_{hh'}$ is much smaller than the total mass $m_h+m_{h'}$ of two hadrons. The error of such an approximation is of $O(\overline{n}^2\Delta)$, where $\overline{n}$ is the average number of particles in $X$ and $\Delta^2=\frac{V_{hh'}}{m_h+m_{h'}}$, see Sec. C in supplemental material. 

\vspace{0.5cm}
\noindent{\it Proof of concept study in NJL model}.- As a proof of concept, we resort to the one-flavor 1+1D NJL model to validate the proposed quantum algorithm. The Lagrangian of the one-flavor NJL model is given by
\begin{equation}\label{LA}
\mathcal{L}=\bar{\psi} (i\gamma^\mu \partial_\mu-m)\psi +g(\bar{\psi} \psi)^2\,,
\end{equation}
where $\psi$ is fermionic operator, $g$ is the strong coupling constant and $m$ is the bare quark mass. The NJL model describes Dirac fermion interactions that can mimic the behavior of QCD to some degree, and capture key characteristics of FFs \cite{Matevosyan:2011ey,Yang:2016gnd,Yang:2020wvt}. In our calculation, the NJL model is discretized with staggered fermion scheme and mapped to qubits using the standard Jordan-Winger transformation \cite{Jordan:1928wi,Fradkin:1989yrd}. The details as well as the qubit Hamiltonian can be found in Ref. \cite{Li:2021kcs}. The qubit representation of quark FF in the hadron rest frame can be written as 
\bea\label{eq:FTFF}
&    zD^h_i(z)=\sum_y e^{-iy m_h/z} \Tilde{D}^h_i(y)\,,
\eea
with
\bea
\Tilde{D}^h_q(y)=&\sum_{k,l=0}^1 \bra{\Omega}e^{iHy}
        \phi_{-2y+k}
		e^{-iHy}\mathcal{H}
    \phi^\dagger_{l}\ket{\Omega}\,,
\eea
where $\phi_{n}=\prod_{i<n}\sigma^3_{i}\sigma_{n}^+$ is a string of Pauli operators~\cite{backens_shnirman_makhlin_2019}. Implementing the proposed quantum algorithm presented in the previous section, the semi-inclusive hadron operator in the NJL model can be written as, 
\bea    \mathcal{H}=M^{-1}&\sum_{ij=0}^{M-1} U(\theta^*)\Tilde{\mathcal{H}}_i U^\dagger(\theta^*) T^j\,,
\eea
where $\Tilde{\mathcal{H}}_i=\ket{\boldsymbol{I}^{h_{q\bar{q}}}_i}\bra{\boldsymbol{I}^{h_{q\bar{q}}}_i}=(I-\sigma^z_{2i})(I+\sigma^z_{2i+1})$. $U(\theta^*)$ can be realized by the variational quantum algorithm and it should map the bare vacuum and one-particle state to the corresponding physical states. In the end, the discretized FFs in position space can be written as
\bea\label{eq:posFF}
    \Tilde{D}^h_q(y)=&M^{-1}\sum_{kl=0}^1 \sum_{ij=0}^{M-1}\bra{\Omega}e^{-iHy}\phi_{-2y+k}e^{iHy}\nn
&\times U(\theta^*) \Tilde{\mathcal{H}}_i U^\dagger(\theta^*)\phi^\dagger_{2j+l}\ket{\Omega}\,, 
\eea
which resembles multiple-point correlation functions. 

We show in Fig.~\ref{fig-qcircuit}(c) the quantum circuit for evaluating Eq.~(\ref{eq:posFF}). With the quantum alternating operator ansatz (QAOA) ~\cite{farhi2014quantum,wiersema-PRXQuantum2020} for constructing $U(\theta^*)$, the circuit depth for preparing vacuum $\ket{\Omega}$ is expected to be $O(N)$. The circuit depth for the controlled $U(\theta^*)$ is $O(N^2)$. According to Ref.~\cite{Li:2021kcs}, the time complexity for Trotter decomposition of $e^{-iHt}$ with an evolution time $t=O(N)$ is $O(N^2/\varepsilon)$ for a desired accuracy $\varepsilon$. The total circuit depth for the circuit in Fig.~\ref{fig-qcircuit} is $O(N^3/\varepsilon)$. The polynomial scaling with the number of qubits shows a quantum advantage of calculating FFs on a quantum computer.

\begin{center}
\begin{figure}[ht]
\centering
\includegraphics[width=0.48\textwidth]{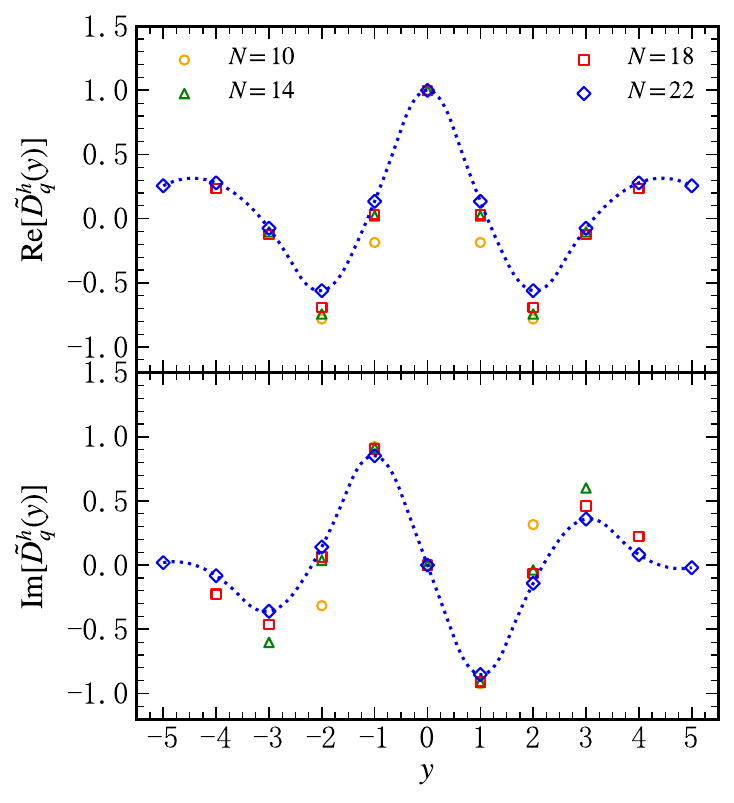}
	\caption{The real (top panel) and imaginary (bottom panel) part of FF in position space with difference qubit number $N$ in NJL model, the dressed quark mass is fixed as $m_q a=0.4$, and hadron mass is $m_h a =0.6$.}
	\label{fig-FFN_pos}
\end{figure}
\end{center}

Using open source package QuSpin \cite{quspin} and projectQ \cite{Steiger2018projectqopensource}, we simulate the quantum circuit in Fig.~\ref{fig-qcircuit} (c) on a classical device with one CPU and one GPU for different system size. We present our results of quark FF using different numbers of qubits, i.e. $N=10$, $N=14$, $N=18$ and $N=22$, with fixed dressed quark mass $m_q a=0.4$ and hadron mass $m_h a=0.6$. Shown in Fig. \ref{fig-FFN_pos} are the real and imaginary parts of FFs in position space, with $\Tilde{D}^h_q$ normalized to ${\rm Re}[\Tilde{D}^h_q(0)]=1$. We find out that $\Tilde{D}^h_q(y)$ converges with the increase of qubit number $N$, which suggests that the contribution from high multiplicity events is suppressed in our considered setup. 

\begin{figure}
    \centering
    \includegraphics[width=0.48\textwidth]{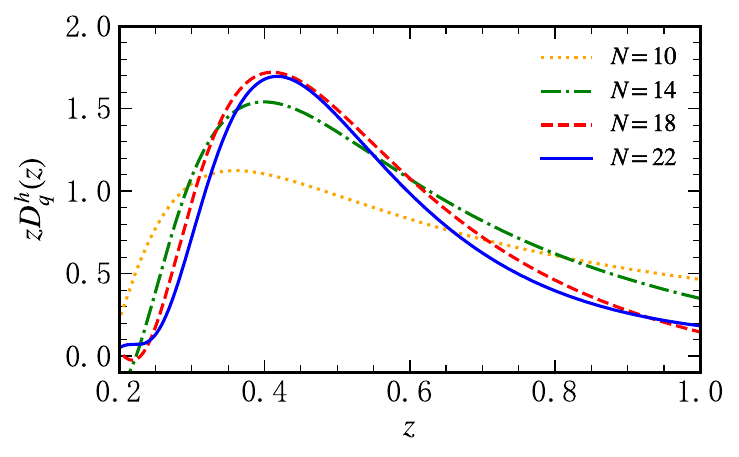}
    \caption{FF in NJL model with different quantum number $N$, which are obtained by Fourier transform of $\Tilde{D}^h_q(y)$ in Fig.~\ref{fig-FFN_pos}.}
    \label{fig-FFN}
\end{figure}

The $N$ dependence of quark FF $D^h_q(z)$ in momentum space is shown in Fig.~\ref{fig-FFN}. The convergence of $D^h_q(z)$ in the small $z$ region is better than that in the large $z$ region, this is because the FF is more significantly affected by the finite volume effect in large $z$ region than that in the small $z$ region. Such finite volume effect also causes the non-zero value $D^h_q(z=1)$, similar to those for PDF calculation in lattice QCD \cite{Ishikawa:2017iym}. 
The shapes shown in Fig.~\ref{fig-FFN} are in qualitative agreement with the extracted quark FFs \cite{Gao:2024nkz} and the calculated charm fragmentation functions using NJL model \cite{Yang:2020wvt} and 1+1D QCD in large $N_c$ limit \cite{Jia:2023kiq}, such as the vanishing of $D^h_q(z)$ in small $z$ region and the peak in large $z$ region.

Moreover, we investigate the effect from quantum noises and explore how error mitigation can improve the results. In the simulation, we adopt a depolarization model to describe noises on quantum hardware. This noisy channel maps the one-qubit density matrix $\rho$ to $\epsilon(\rho)=(1-p)\rho+\sum_i \frac{p}{3}\sigma^i \rho \sigma^i$ after acting arbitrary quantum gates on the qubit. The noise level is set as $p=10^{-3}$ which is feasible for near-term quantum processors. We find out that the effect from quantum noises can not be neglected and could lead to unreliable results, which is caused by the fact that both the vacuum and hadron states prepared with noisy quantum circuits could be mixed with non-physical states of single quark or antiquark. Remarkably, the mixing with non-physical states can make a direct zero-noise extrapolation method~\cite{PhysRevLett.119.180509,9259940} fails to work. We find that it is necessary to first mitigate the noise causing non-physical states by postselection with a projective measurement on the auxiliary qubit, see Sec. D in supplemetal material. We incorporate postselection for FFs in position space with higher-order Richardson extrapolation method \cite{PhysRevLett.119.180509}
\bea
    \Tilde{D}^{h,\lambda}_q(y) = \sum_{j=0}^\lambda \gamma_j \Tilde{D}^h_q(y;c_jp)\,,
\eea
where $\gamma_j = \prod_{m\not=j} c_m (c_j-c_m)^{-1}$, $\Tilde{D}^{h,\lambda}_q(y)$ is the $\lambda$-th order error mitigation result of noisy FF $\Tilde{D}^h_q(y;c_jp)$, $c_j$ is free scale parameters. Here, we fix $p=10^{-3}$ and vary $c_j=(1+0.1j)\times 10^{-3},\ j=1,2,...,7$.
\begin{figure}[htbp]
	\centering
\includegraphics[width=0.48 \textwidth]{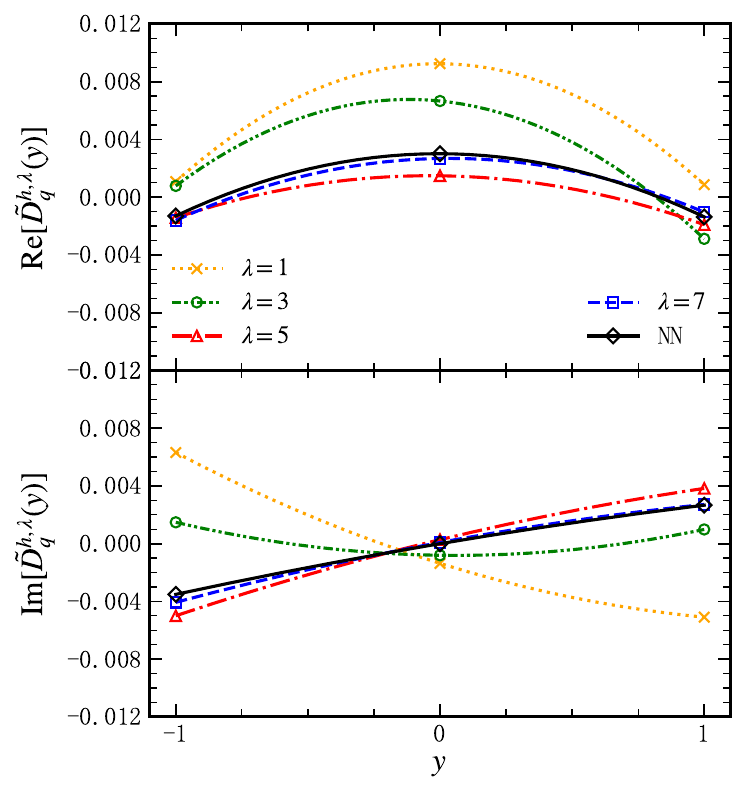}
	\caption{Real (top panel) and imaginary (bottom panel) part of FF in position space with $N=6$, $m a=1$, $g =0.8$ and quantum gate fidelity $0.999$, NN is the result with the zero-noise quantum circuit, $\lambda=1$ is the noisy result after postselection and $\lambda=k\,,k>1$ is the $k$-th Richardson extrapolation result after postselection.}
\label{fig-FFmit}
\end{figure}
As shown in Fig.~\ref{fig-FFmit}, both $\Re[\Tilde{D}^h_q(y)]$ and $\Im[\Tilde{D}^h_q(y)]$ converge to the zero-noise results with higher-order Richardson extrapolation method, indicating the effectiveness of error mitigation in our proposed algorithm for direct calculation of FFs.

\vspace{0.5cm}
\noindent{\it Summary}.- We proposed a framework for the first direct calculation of FFs utilizing quantum computing methodology. In particular, we constructed the hadron operator for filtering out semi-inclusive hadrons of desired types in a collection of multi-particles encoded in the quantum state. As a proof of concept, we demonstrated the viability of our proposed quantum computing framework by calculating quark FF in the 1+1 dimensional NJL model up to $22$ qubits. Moreover, we show explicitly that it is necessary to incorporate the postselection of physical states into the error mitigation for accurately determining FFs on noisy quantum processors. Our exploratory study manifests the quantum advantage in the direct calculation of FFs, which paves the way for investigating fragmentation on near-term quantum hardwares.  

\vspace{0.5cm}
\noindent{\it Acknowledgements}.- We thank Xiaohui Liu, Xingyu Guo and Wai Kin Lai for helpful discussions. The work is supported by the Guangdong Major Project of Basic and Applied Basic Research No.~2020B0301030008 and 2023A1515011460, and NSFC under Grants Nos.~12035007, 12375013 and 12022512.

%%%%%%%%%%%%%%%%%%%%%%%%%%%%%%%%%%%%%%%%%%%%%%%%%%%%%%%%%%%%
%\begin{thebibliography}{99}

%\end{thebibliography}
\normalem
%\bibliographystyle{h-physrev5}   
%\bibliography{refs.bib}

\section*{Supplemental Material}

\subsection*{A. Construction of semi-inclusive hadron operator}
The construction of semi-inclusive hadron operator (SIHO) has been presented in the main text
\bea
    \mathcal{H} &= \left(\sum_{i=0}^{M-1} U\ket{\boldsymbol{I}^h_i}\bra{\boldsymbol{I}^h_i}U^\dagger\right) \times \left(\sum_{j=0}^{M-1} \frac{1}{M} T^j\right)\nn
    &\equiv \mathcal{N}_h \times P(\mathbf{p}=0)\,.
\eea
In the following part of this section, we will show that $\mathcal{N}_h$ and $P(\mathbf{p}=0)$ can be regarded as number operators of hadron $h$ and zero-momentum projector, respectively. 

We can act $\mathcal{N}_h$ on multi-particle states to explain why $\mathcal{N}_h$ is a number operator of hadrons. A multi-particle state with $n_h$ hadrons of $h$ localized in positions $i_1,i_2,..,i_{n_h}$ can be denoted as $\ket{h_{i_1},h_{i_2},...,h_{i_{n_h}},\Psi_{\{j\not=i_1,i_2,..,i_{n_h}\}}}$, where other particles $\Psi\not=h$ and they distribute at the positions except $i_1,i_2,...,i_{n_h}$. Note that $\mathcal{N}_h$ commutes with $U$ as $U$ is quantum-number preserving, then we have
\bea
    &\quad \ \mathcal{N}_h \ket{h_{i_1},h_{i_2},...,h_{i_{n_h}},\Psi_{\{j\not=i_1,..,i_{n_h}\}}} \nn
    &=\sum_{i=0}^{M-1} U \ket{\boldsymbol{I}^h_i}\bra{\boldsymbol{I}^h_i} \otimes {\rm Id}_{\{k\not=i\}} \ket{\boldsymbol{I}^h_{i_1},\boldsymbol{I}^h_{i_2},...,\boldsymbol{I}^h_{i_{n_h}},\boldsymbol{I}^\Psi_{\{j\not=i_1,..,i_{n_h}\}}}\nn
    &=U n_h \ket{\boldsymbol{I}^h_{i_1},\boldsymbol{I}^h_{i_2},...,\boldsymbol{I}^h_{i_{n_h}},\boldsymbol{I}^\Psi_{\{j\not=i_1,..,i_{n_h}\}}}\nn
    &=n_h \ket{h_{i_1},h_{i_2},...,h_{i_{n_h}},\Psi_{\{j\not=i_1,..,i_{n_h}\}}}\,.
\eea
The above equation tells us that the operator $\mathcal{N}_h$ can count the number of hadrons $h$ in multi-particle states.

As mentioned in the main text, the FF needs to be calculated in the hadron rest frame. In this frame, the state $\ket{h,X}$ also has zero momentum because we focus on collinear FFs, which have $p_h+p_X = p_h/z$. We need a zero-momentum projector to guarantee the operator $\mathcal{N}_h$ counting the zero-momentum hadron. To verify $\Tilde{P}(\mathbf{p}=0)$ is such operator, we calculate  $\Tilde{P}({\boldsymbol{p}}=0)^2$ at first,
\bea
    \Tilde{P}({\boldsymbol{p}}=0)^2 &= \frac{1}{M^2} \sum_{ij=1}^{M} T^i T^j\nn
    & = \frac{1}{M^2}\sum_{i=1}^{M} M T^i = \Tilde{P}({\boldsymbol{p}}=0)\,.
\eea
showing $\Tilde{P}({\boldsymbol{p}}=0)$ is a projector. Then we show that $\Tilde{P}({\boldsymbol{p}}=0)$ acts on a zero-momentum state will keep it unchanged while acts on the non-zero-momentum state will vanish,
\bea\label{eq-propz}
    \Tilde{P}({\boldsymbol{p}}=0)\ket{\boldsymbol{p} = 0} = \frac{1}{M}\sum_{j=1}^{M-1} T^j \ket{\boldsymbol{p} = 0} = \ket{\boldsymbol{p} = 0}\,.
\eea
The last step in Eq.~(\ref{eq-propz}) is because $T\ket{\boldsymbol{p} = 0} = \ket{\boldsymbol{p} = 0}$.
\bea\label{eq-prop}
    \Tilde{P}({\boldsymbol{p}}=0)\ket{\boldsymbol{p}} = \frac{1}{M}\sum_{j=1}^{M-1} T^j \ket{\boldsymbol{p}} = \frac{1}{M} \sum_{j=1}^{M-1} e^{i p_k j}\ket{\boldsymbol{p}} = 0\,.
\eea
The last step in Eq.~(\ref{eq-prop}) is because $\sum_{j=0}^{M-1} e^{i p_k j} = 0$, where $p_k = \frac{2\pi k}{M}\,,k=1,...,M-1$ on one dimensional lattice. In conclusion, the $\Tilde{P}({\boldsymbol{p}}=0)$ is exactly the zero-momentum projector.

\subsection*{B. Variational quantum algorithm to construct $U$}
\label{sec:VQA}
The unitary operator $U$ can be constructed by a variational quantum algorithm (VQA) \cite{Li:2021kcs}. For our purpose, we require $U$ to prepare the vacuum and one-hadron states instead of preparing all multiple hadron states.  Such $U$ suffices for transforming the computational basis of multiple bare hadrons to states of multiple physical hadrons, once energy is dominated by the mass of physical hadrons while their interaction energy can be comparably very small, as justified later in Sec. C~\ref{sec:Error_SIHO}. For demonstration, we use the NJL model to show how the VQA works. The qubit Hamiltonian of the NJL model can be found in \cite{Li:2021kcs}, which can be written as $H = \sum_{j=1}^4 H_j$. Every $H_j$ commutes with all the conservation charges of the NJL model hence the time evolution $\exp(-i H_j t)$ preserves the quantum number of the NJL model, so the variational ansatz $U$ can be written as
\bea
    U(\theta) = \prod_{i=1}^{p} \prod_{j=1}^{4} \exp(i \theta_{ij} H_j)\,.
\eea
The desired $U(\theta)$ needs to map the corresponding four computational basis to vacuum $\ket{\Omega}$, anti-quark state $\ket{\bar{q}}$, quark state $\ket{q}$ and $q\bar{q}$ bound state $\ket{h_{q\bar{q}}}$. Those four computational basis are denoted as $\ket{\Tilde{\Omega}}$, $\ket{\Tilde{\bar{q}}}$, $\ket{\Tilde{q}}$ and 
$\ket{\Tilde{h}_{q\bar{q}}}$
\bea\label{eq-refs}
    &\ket{\Tilde{\Omega}} = \ket{\boldsymbol{I}^\Omega_0,...,\boldsymbol{I}^\Omega_{M-1}}\,,\nn
    &\ket{\Tilde{\bar{q}}} = \frac{1}{M}\sum_{i=0}^{M-1}\ket{\boldsymbol{I}^\Omega_0,\boldsymbol{I}^{\Omega}_1,...,\boldsymbol{I}^{\bar{q}}_{i},\boldsymbol{I}^{\Omega}_{i+1},...,\boldsymbol{I}^{\Omega}_{M-1}}\,,\nn
    &\ket{\Tilde{q}} = \frac{1}{M}\sum_{i=0}^{M-1}\ket{\boldsymbol{I}^\Omega_0,\boldsymbol{I}^{\Omega}_1,...,\boldsymbol{I}^q_{i},\boldsymbol{I}^{\Omega}_{i+1},...,\boldsymbol{I}^{\Omega}_{M-1}}\,,\nn
    &\ket{\Tilde{h}_{q\bar{q}}} = \frac{1}{M}\sum_{i=0}^{M-1}\ket{\boldsymbol{I}^\Omega_0,\boldsymbol{I}^{\Omega}_1,...,\boldsymbol{I}^{h_{q\bar{q}}}_{i},\boldsymbol{I}^{\Omega}_{i+1},...,\boldsymbol{I}^{\Omega}_{M-1}}\,,
\eea
where $\ket{\boldsymbol{I}^\Omega}=\ket{01}$, $\ket{\boldsymbol{I}^{\bar{q}}}=\ket{00}$,
$\ket{\boldsymbol{I}^q}=\ket{11}$ and $\ket{\boldsymbol{I}^{h_{q\bar{q}}}}=\ket{10}$ in the NJL model. All four states in Eq.~(\ref{eq-refs}) are Dicke-like states, which can be prepared by the quantum circuit in \cite{Li:2022lyt}. To obtain the desired $U(\theta)$ we need to optimize parameters $\theta$ by minimizing the loss function
\bea \label{eq:obj}
    E(\theta) =& w_\Omega \braket{\Tilde{\Omega}|U^\dagger(\theta) H U(\theta)|\Tilde{\Omega}}\nn
    &+\sum_{\Psi} w_\Psi \braket{\Tilde{\Psi}|U^\dagger(\theta) H U(\theta)|\Tilde{\Psi}}
\eea
where $\Psi=q,\bar{q},h_{q\bar{q}}$ and 
$w_i$ could be chosen freely so long as $w_\Omega>w_{\bar{q}}=w_q>w_{h_{q\bar{q}}}>0$. $w_{\bar{q}}=w_q$ is due to energies of $q$ and $\bar{q}$ should be the same.
eThe optimized parameters are denoted as $\theta^*$, then we have $\ket{\Omega} = U(\theta^*)\ket{\Tilde{\Omega}}$, $\ket{\bar{q}} = U(\theta^*)\ket{\Tilde{\bar{q}}}$, $\ket{q} = U(\theta^*)\ket{\Tilde{q}}$, $\ket{h_{q\bar{q}}}=U(\theta^*)\ket{\Tilde{h}_{q\bar{q}}}$. Now, in the NJL model, the operator $U$ can be approximated by $U(\theta^*)$, because it maps the computational basis to vacuum and one-particle states and preserves all the symmetry of the NJL model.

\subsection*{C. Error estimation of SIHO}
\label{sec:Error_SIHO}
In the VQA method, the operator $U$ can only prepare vacuum and one-hadron state, the multi-hadron states are prepared approximately. So the error of the SIHO needs to be estimated. We assume the two-hadron state prepared by $U$ is $\ket{\bar{h}_\alpha,\bar{h}_\beta}$ and the exact two-hadron state is $\ket{h_\alpha,h_\beta}$, which is an eigenstate of Hamiltonian $H$. Obviously, the state $\ket{\bar{h}_\alpha,\bar{h}_\beta}$ can not reflect the interaction between two hadrons, so we have
\bea\label{Eq:Heps}
    \braket{\bar{h}_\alpha,\bar{h}_\beta|H|\bar{h}_\alpha,\bar{h}_\beta} &= \braket{h_\alpha,h_\beta|H|h_\alpha,h_\beta}+\epsilon_{\alpha\beta} \nn
    &= E_{\alpha\beta}+\epsilon_{\alpha\beta}\,,
\eea
where $E_{\alpha\beta}$ is the 
energy eigenvalue of the state $\ket{h_\alpha,h_\beta}$. The energy deviation $\epsilon_{\alpha\beta}$ is roughly equal to the interaction potential between two hadrons, which is denoted as $V_{h_\alpha h_\beta}$. It should be stressed $V_{h_\alpha h_\beta}$ is not relevant to the original interaction of the Hamiltonian. Rather, it is residual interaction. The state $\ket{\bar{h}_\alpha,\bar{h}_\beta}$ can be decomposed as
\bea\label{Eq:hdel}
    \ket{\bar{h}_\alpha,\bar{h}_\beta} = \frac{1}{\sqrt{1+\Delta_{\alpha\beta}^2}}\ket{h_\alpha,h_\beta}+\frac{\Delta_{\alpha\beta}}{\sqrt{1+\Delta_{\alpha\beta}^2}}\ket{\varphi_{\alpha\beta}}\,,
\eea
where $\ket{\varphi_{\alpha\beta}}$ is orthogonal to $\ket{h_\alpha,h_\beta}$. Plug Eq.~(\ref{Eq:hdel}) into LHS of Eq.~(\ref{Eq:Heps}) and ignore the $O(\Delta_{\alpha\beta}^2)$ term, $\Delta_{\alpha\beta}$ can be written as
\bea
    \Delta_{\alpha\beta} = \sqrt{\frac{\abs{\epsilon_{\alpha\beta}}}{E_{\alpha\beta}}} \approx \sqrt{\frac{\abs{V_{h_\alpha h_\beta}}}{m_{h_\alpha}+m_{h_\beta}}}\,.
\eea
The SIHO needs to sum over all multi-hadron states. For the $n$-hadron state, it has approximately $n^2$ two-hadron pairs. So the error of the SIHO $\epsilon_h$ can be written as
\bea
    \epsilon_h = O[\bar{n}^2\, {\rm max}({\Delta_{\alpha\beta}})]\equiv O(\bar{n}^2 \Delta),
\eea
where $\bar{n}$ is the average number of particles in $X$ and ${\rm max}({\Delta_{\alpha\beta}})$ is the maximum of the  $\Delta_{\alpha\beta}$ for all $\alpha$ and $\beta$. It can be expected that the mass of a hadron stems mainly from the original interaction of quarks, while the residue interaction between hadrons is negligibly small. Thus, for a fixed $\bar{n}$, the error $\epsilon_{h}$ of the SIHO is small because the hadron mass $m_{h_\alpha}+m_{h_\beta}$ will be several orders of magnitude larger than the interaction potential between hadrons.

\subsection*{D. Error mitigation of FFs}
The depolarization noise channel is used to simulate the noisy quantum computer. The noise channel maps the noiseless density matrix $\rho$ to $\epsilon(\rho)$
\bea
    \epsilon(\rho) = (1-p)\rho+\frac{p}{3}(\sigma^x \rho \sigma^x+\sigma^y \rho \sigma^y+\sigma^z \rho \sigma^z),
\eea
where $1-p$ is the noise parameter that reflects the quantum gate fidelity. We set $p\approx 10^{-3}$, which is reasonable for Noisy Intermediate-Scale Quantum (NISQ) devices. For a $N=6$ qubits quantum computer, the depth of the quantum circuit to simulate FFs is about $O(10^2)$. Due to the existence of SIHO, FF needs to be calculated by adding up $(N/2)^2 \approx  10$ output results from different quantum circuits. The noise effect may accumulate after adding up those outputs, so the noise effect can not be neglected. To obtain reasonable output from quantum computing, an error mitigation algorithm is needed. In the main text, the error mitigation algorithm is made up of two parts, where the first part is postselection and the second part is Richardson zero noise extrapolation.
\begin{figure*}[htbp]
    \centering
    \includegraphics[width=0.98\textwidth]{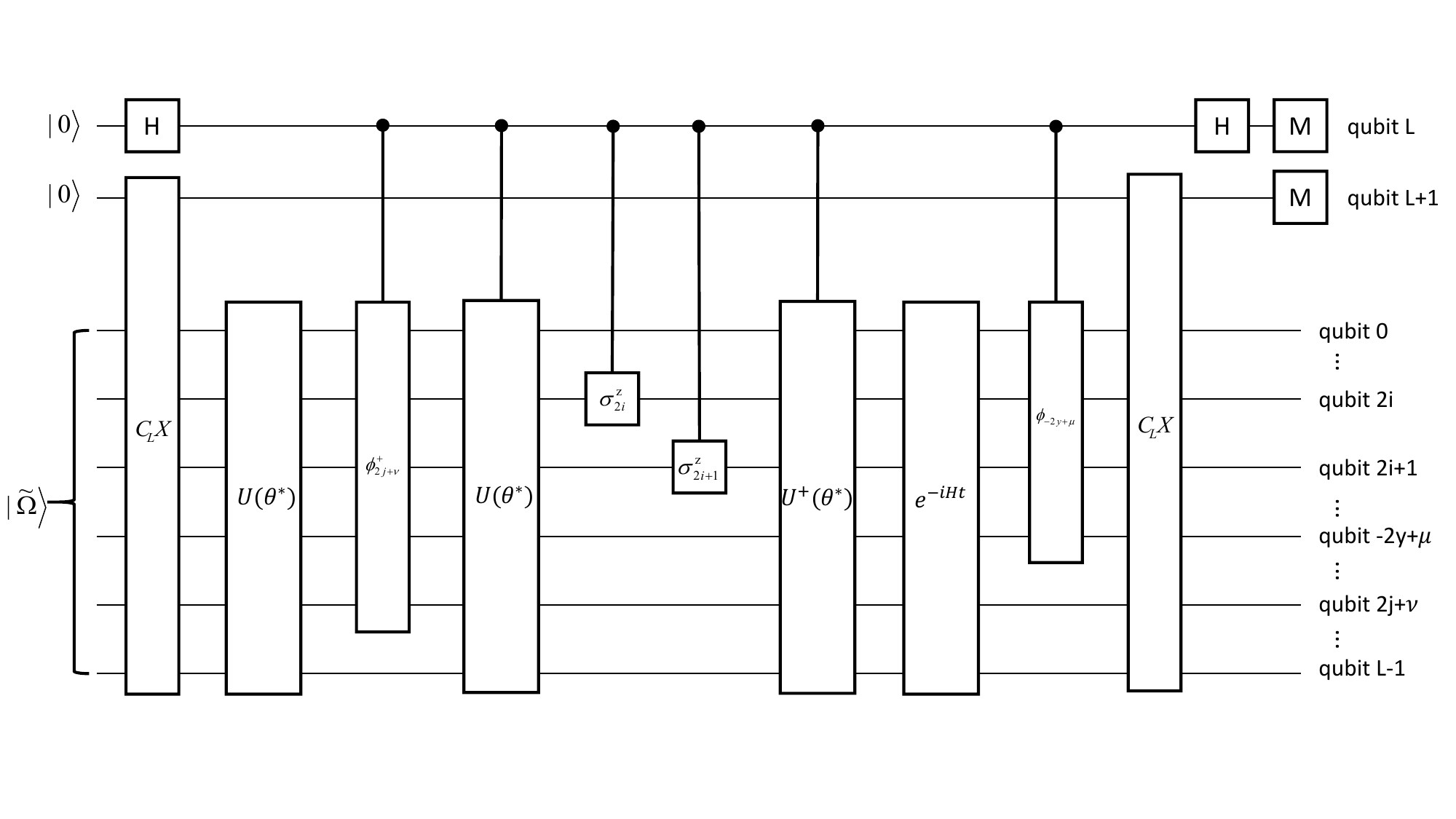}
    \caption{Quantum circuit for postselection.}
    \label{fig-post}
\end{figure*}

\begin{figure}[htbp]
    \centering
    \includegraphics[width=0.98\linewidth]{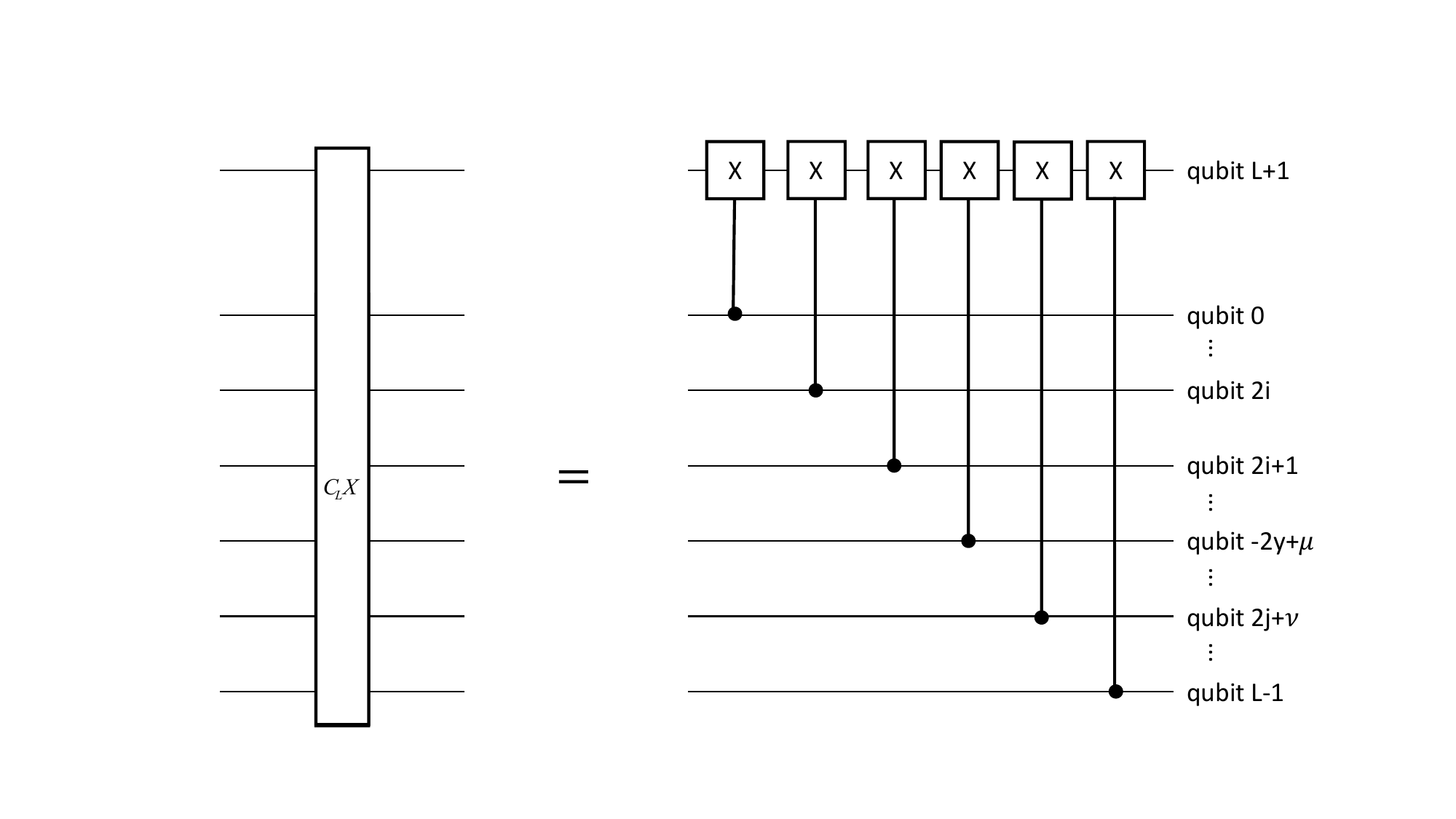}
    \caption{Definition of gate $C_L X$.}
    \label{fig-clx}
\end{figure}

\begin{figure}[htbp]
    \centering
    \includegraphics[width=0.98\linewidth]{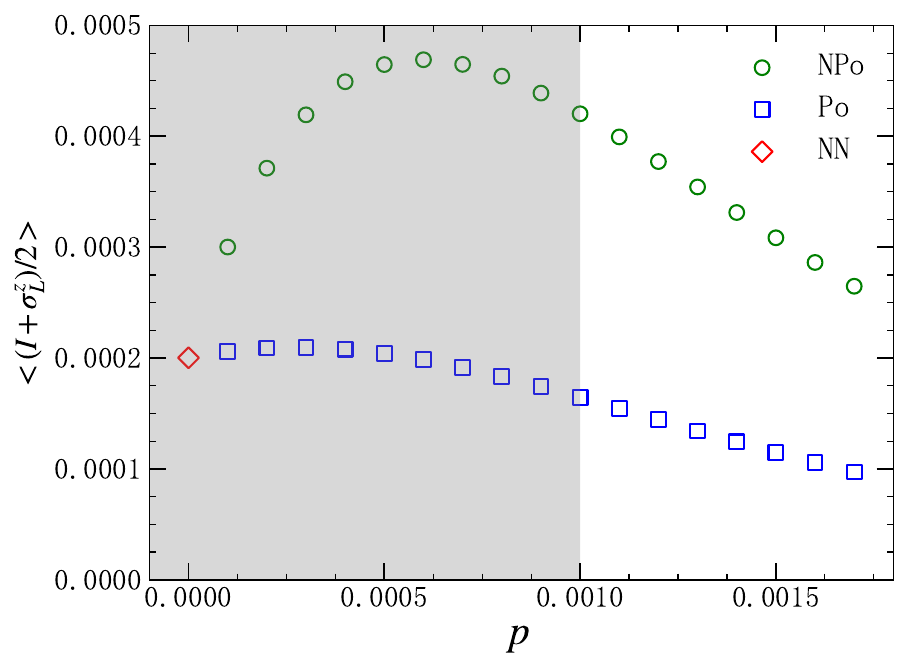}
    \caption{Results of postselection and no postselection with different noise parameters $p$. The NISQ device can only obtain the result with $p>10^{-3}$ (outside of the shadow region).}
    \label{fig-ponpo}
\end{figure}

In the main text, we point out that the unitary operator $U$ need to protect the quantum number. However, the $\sigma^x$ and $\sigma^y$ errors will change the charge of a quantum state by flipping the qubit. We chose to select the measurement results with even numbers of $\sigma^x$ and $\sigma^y$ errors in the quantum circuit. The above procedure is called postselection. Still, it is possible that, for instance, a $q$ state will become a $qqq$ state due to noise, which can not be detected in the postselection. However, we expect that for small $p$ the one-flip process dominates. In this regard, the postselection can restore the charge conservation in a good approximation. Fig.~\ref{fig-post} shows the postselection quantum circuit, where the definition of $C_L X$ gate is shown in Fig.~\ref{fig-clx}. We select measurement results with $\sigma^z_{L+1}=+1$ to calculate the FF because the odd numbers $\sigma^x$ and $\sigma^y$ error will flip the qubit $L+1$ from $\ket{0}$ to $\ket{1}$.  In Fig.~\ref{fig-ponpo}, we show how the measurement result of qubit $L$ depends on the noise parameter $p$,  where NPo is no postselection result, Po is the postselection result and NN is the no-noise result. The NISQ device can not reach the shadow area in Fig.~\ref{fig-ponpo} because the noise of this area is smaller than $10^{-3}$. From Fig.~\ref{fig-ponpo}, it can be seen that extrapolating zero noise based on the data in the range of $p>10^{-3}$ for NPo can easily lead to incorrect results. Meanwhile, the extrapolation of Po data can give us correctly no noise result, which shows the effectiveness of postselection. 

There are still some errors after postselection. Those errors can be mitigated by Richardson zero noise extrapolation \cite{PhysRevLett.119.180509}. Richardson extrapolation suggests one can run a quantum circuit under different noise parameters, then the error can be mitigated by extrapolating the result to zero noise by
\bea
    O^\lambda(p) = \sum_{j=0}^\lambda \gamma_j O(c_j p)
\eea
where $\gamma_j = \prod_{m\not=j} c_m (c_j-c_m)^{-1}$, $O^\lambda(p)$ is the $\lambda$-th order error mitigation result of observable $O$ and $c_j$ is the scale parameters which can be chosen freely.

\end{document}